\documentclass[12pt]{article}
\usepackage{amsmath}
\usepackage{amsxtra}
\usepackage{amsthm}
\usepackage{amssymb}     
\usepackage{graphicx}
\usepackage[all]{xy}
\newtheorem*{prop}{Proposition}
\newtheorem*{defin}{Definition}
\newtheorem*{remark}{Remark}

\def\id{{\text{1} \kern-.26em \text{l}}}

\begin{document}
\vspace*{1cm}
\begin{center}{\large \bf On the ubiquity of matrix-product states in
    one-dimensional stochastic processes with boundary interactions}\\
\vspace{5mm}{\bf Kai Klauck\footnote{e-mail: {\tt kok@thp-uni-koeln.de}}, 
Andreas Schadschneider\footnote{e-mail: {\tt as@thp-uni-koeln.de}}}\\
{\it Institut f\"ur Theoretische Physik,
Universit\"at zu K\"oln, D-50937 K\"oln, Germany}\\
\end{center}
\begin{center}
\today
\end{center}
\vspace{10mm} {{\bf Abstract.} Recently it has been shown that the
  zero-energy eigenstate -- corresponding to the stationary state -- of a
  stochastic Hamiltonian with nearest-neighbour interaction in the bulk and
  single-site boundary terms, can always be written in the form of a so-called
  matrix-product state. We generalize this result to stochastic
  Hamiltonians with arbitrary, but finite, interaction range. 
As an application two different particle-hopping models with three-site
  bulk interaction are studied. For these models which can be interpreted
as cellular automata for traffic flow, we present exact solutions for periodic
  boundary conditions and some suitably chosen  boundary interactions.
}

\hyphenation{one-dimen-sional}

\newpage


\section{Introduction}
The so-called Matrix-Product Ansatz (MPA) was
developed and used for the construction of Optimum Ground States (OGS) in
quantum spin chains (see e.g.\ \cite{zitt1,zitt2,mikeska} and references
therein). Soon the method found its way into the field of one-dimensional 
stochastic processes. The most prominent
example in this area  is the solution of the totally asymmetric simple
exclusion process (TASEP) with open
boundaries by Derrida, Evans, Hakim and Pasquier (DEHP)
\cite{derrida}~\footnote{At the same time a solution using a different
approach has been presented in \cite{domany-schuetz}.}, which has reached a
paradigmatic status in the field of low-dimensional nonequilibrium
processes.   
\newline
The general principle behind the MPA for stationary states of
stochastic Hamiltonians -- the so-called \emph{cancelling-mechanism} -- 
was recognized  by Hinrichsen {\it et al} \cite{hinrichsen},
allowing the treatment of more complex reaction-diffusion systems.
The cancelling-mechanism may be considered as a generalization
of the OGS-concept to stochastic systems \cite{rsss}.
\newline
Based on this cancelling-mechanism, Krebs and Sandow (KS) \cite{ks}
could prove that for stochastic processes in one dimension with
nearest-neighbour interaction in the bulk and boundary fields acting 
only on the first and the last site, the MPA is not an Ansatz, but merely 
a reformulation of the problem, i.e.\ the stationary state of such
processes can always be written as a matrix-product state (MPS).
This means that -- in contrast to the situation for ground states of
quantum systems -- (generalized) OGS are generic for stationary 
states of stochastic models.

Many stochastic systems arising in the context of such fields as
traffic flow, granular matter, chemical reactions and biological
motion  have naturally an
interaction, which is not restricted to nearest
neighbours. However, the only example of an application of the MPA to models
with an interaction range $r \geq 2$ seems to be the work of
E\ss ler and Rittenberg \cite{essler}. These authors introduced a three-site
($r=3$) version of the cancelling-mechanism used implicitely in \cite{derrida}.
Preliminary studies performed by us indicated, however, that this mechanism
is not the most general one and allows only the solution of
some special models. Therefore it was natural to look for a generalization
of the proposition by KS in order to identify the most general mechanism
for arbitrary interaction range $r\geq 2$.

Indeed we found that this generalization is possible.
In this letter we want to show how the mechanism of KS has to be modified
to be amenable to systems with interaction range $r\geq 2$.
After clarifying the cancelling-mechanism -- which is the main result of
this letter -- the proof of our propositions 
is a straightforward generalization of that of KS. We therefore omit it
here and postpone the details to a  later publication
\cite{spaeta}. Moreover we want to show the usefulness of our method,
by solving two different particle-hopping models with three-site interaction  
in the bulk for  periodic boundary conditions and some suitably chosen
boundary interactions. Both models are interesting on their own since they
can be interpreted as cellular automata describing traffic flow on a
highway.


\section{The generalized Krebs-Sandow proposition}

We split the original proposition of Krebs and Sandow in two independent 
propositions A and B. The main reason is that proposition A gives a 
sufficient local criterion for the existence of MPS and covers stochastic 
processes both with boundary interactions and periodic boundary 
conditions. Proposition B deals with boundary interactions only where it
is possible to show explicitly that the criterion of proposition A can 
always be fulfilled.

Consider a stochastic process on a chain consisting of
$L$ sites, where each site can be in one of $m$ states. For processes in
continuous time the temporal evolution of the probability
vector is governed by the master equation~\footnote{For an introduction 
to master equations in form of an imaginary-time Schr\"odinger equation,
see \cite{rittenberg-droz} and references therein.}
\begin{equation}\label{eq:master}
  \partial_t|P_L(t)\rangle=-\hat{H}_L|P_L(t)\rangle.
\end{equation}
The linear operator $ \hat{H}_L$ is a stochastic Hamiltonian of the form
\begin{equation}
  \Hat{H}_L(r)=\hat{h}_{{\rm left}}(r)+\sum_{k=1}^{L-r+1}
\hat{h}_{k,k+1,\ldots,k+r-1}+\hat{h}_{{\rm right}}(r),
\label{eq:Hdefopen}
\end{equation}
in the case of boundary interactions. 
\newline
For periodic boundary conditions (p.b.c.) it has the form
\begin{equation}
  \Hat{H}^{(p)}_L(r)=\sum_{k=1}^{L}\hat{h}_{k,k+1,\ldots,k+r-1}.
\label{eq:Hdefper}
\end{equation} 
The integer $r$ denotes the range of
interaction in the bulk, i.e. $\hat{h}_{k,k+1,\ldots,k+r-1}$ acts on
$r$ sites beginning with site $k$; $\hat{h}_{{\rm left}}(r)$ and
$\hat{h}_{{\rm right}}(r)$ are boundary interactions acting on the first, 
respectively last, $r-1$ sites of the chain. Note that the stochastic 
Hamiltonians (\ref{eq:Hdefopen}) and
(\ref{eq:Hdefper}) are in general non-hermitian.

From now on we will only be concerned with the steady state solution
of (\ref{eq:master}), which is the  eigenvector
$|P_L\rangle_0$ of $\hat{H}_L$ with eigenvalue 0. Since $\hat{H}_L$ 
is a stochastic matrix, at least one such eigenvector 
exists \cite{vankampen}.
We will refer to $|P_L\rangle_0$ as zero-energy eigenvector
in the following.
\newline
Moreover we address the question
whether   $|P_L\rangle_0$ can be written as a matrix-product state
\begin{equation}
|P_L\rangle_0=\frac{1}{Z_L}~\langle W| \mathcal{D}^{\otimes L}|V \rangle,
\end{equation}
in the presence of boundary interactions. 
\newline
For periodic boundary conditions one has to modify the ansatz to guarantee
translational invariance:
\begin{equation}
|P_L\rangle_0=\frac{1}{Z_L}~\operatorname{Trace}[\mathcal{D}^{\otimes L}].
\label{eq:stat_period}
\end{equation}
In both cases $\mathcal{D}$ is a vector of dimension $m$ with components
$D_i$, where the $D_i$ are matrices acting on some auxiliary vector
space $A$. $|V \rangle\in A$ and $\langle W|\in A^*$ are vectors in $A$ 
and its dual $A^*$, respectively.

Proposition A gives a sufficient local criterion for the existence of 
a matrix-product state - the so-called \emph{cancelling-mechanism}. 
Before we proceed we have to make the following definition:
\begin{defin}  
  $\mathcal{X}(r)$ is a column-vector with $m^{r-1}$ entries
  $X_{i_1,i_2,\ldots,i_{r-1}}$, where $i_\gamma\in\lbrace
  1,2,\ldots,m\rbrace$. The $X_{i_1,i_2,\ldots,i_{r-1}}$ are matrices
  acting on the vector space $A$ defined above. The position of
  $X_{i_1,i_2,\ldots,i_{r-1}}$ in the vector $\mathcal{X}(r)$ is
  given by $1+(i_1-1)+m(i_2-1)+\ldots+m^{r-2}(i_{r-1}-1)$.
\end{defin}
Using this definition we now can formulate the first part of the
generalized KS-proposition:
\begin{prop}[A]
(i) If one can find $m$ matrices $D_i$ and $m^{r-1}$ matrices
  $\mathcal{X}(r)$ such that they fulfil
\begin{equation}
\label{eq:mechanism-bulk}
\hat{h}_{k,k+1,\ldots,k+r-1}(\overbrace{\mathcal{D}\otimes\mathcal{D}
\otimes\cdots\otimes\mathcal{D}}^{r\text{~times}})
=\mathcal{X}(r)\otimes\mathcal{D}-\mathcal{D}\otimes\mathcal{X}(r)
\end{equation}
then $|P_L\rangle_0=\operatorname{Trace}[\mathcal{D}^{\otimes L}]$ is
an zero-energy eigenvector of $\Hat{H}^{(p)}_L(r)$ (see (\ref{eq:Hdefper})), 
i.e.\ a stationary state of the underlying stochastic process. 
\newline
(ii) If in addition to (\ref{eq:mechanism-bulk}) one can find vectors 
$|V\rangle\in A$ and $\langle W|\in A^*$ such that
\begin{equation}
\langle
    W|\hat{h}_l(r)(\overbrace{\mathcal{D}\otimes\mathcal{D}\otimes\cdots
\otimes\mathcal{D}}^{r-1 \text{
    times}})
=-\langle W|\mathcal{X}(r)\label{eq:mechanism-bound1},
\end{equation}
and
\begin{equation}
\hat{h}_r(r)(\underbrace{\mathcal{D}\otimes\mathcal{D}\otimes\cdots
\otimes\mathcal{D}}_{r-1 \text{
    times}})|V\rangle 
=\mathcal{X}(r)|V\rangle\label{eq:mechanism-bound2},
\end{equation}
then $|P_L\rangle_0=\langle W| \mathcal{D}^{\otimes L}|V \rangle$ is a
zero-energy eigenvector of $\Hat{H}_L(r)$ (see (\ref{eq:Hdefopen}))
with boundary interactions $\hat{h}_{{\rm left}}(r)$ and 
$\hat{h}_{{\rm right}}(r)$.
\end{prop}
\begin{proof}The proof is a straightforward generalization of the
  proof in \cite{ks}. For details see \cite{spaeta}.
\end{proof}
\begin{remark}
The relations (\ref{eq:mechanism-bulk}), (\ref{eq:mechanism-bound1})
and (\ref{eq:mechanism-bound2}) are the most general cancelling-mechanism 
for the stochastic processes considered here.
\end{remark}

In the case of boundary interactions we are able to show more. Here
the matrix-product state is not an ansatz, but merely a
reformulation of the fact that the stationary state is a zero-energy
eigenvector of $\Hat{H}_L(r)$ for all system lengths.
\begin{prop}[B]
Given a stochastic process described by a stochastic Hamiltonian of
the form (\ref{eq:Hdefopen})
which has a unique stationary state for any system length $L$.
Then the eigenstate $|P_L\rangle_0$ with eigenvalue 0 corresponding to 
this stationary state can be written as a 
matrix-product state $\langle W| \mathcal{D}^{\otimes L}|V \rangle$ with 
$\mathcal{D}=(D_1,D_2,\ldots,D_m)^t$ and vectors $\langle W|$, $|V\rangle$. 
Moreover one finds $m^{(r-1)}$  matrices
$\mathcal{X}(r)$, such that the cancelling-mechanism 
(\ref{eq:mechanism-bulk})-(\ref{eq:mechanism-bound2}) is fulfilled.
\end{prop}
\begin{proof}See \cite{spaeta}. As in \cite{ks} one can give an
explicit construction for the operators involved.
\end{proof}
\begin{remark}
In the case of $m=2$ and $r=3$ it is easy to see, that the
cancelling-mechanism proposed in \cite{essler} is a special case of
our mechanism. Using this special mechanism a proposition similar to
our proposition A could be formulated. Nevertheless our
mechanism has the advantage that in case of boundary interactions it
leads to an algebra which is manifestly non-trivial~\footnote{By a
trivial algebra we mean an algebra which is equal to 0.}. This stems
from the fact that in the course of the proof of proposition B we
construct an explicit non-trivial representation of this algebra. A
discussion of this point can be found in \cite{spaeta}.  
\end{remark}


\section{Application of the proposition}
We present now applications of our cancelling-mechanism by determining
the stationary states of one-dimensional two-state ($m=2$) 
reaction-diffusion systems with three-site interactions in the
bulk. In the following we investigate two models which are 
generalizations of the TASEP.  
\newline
In the first model (\emph{model A})  particles hop exclusively in one 
direction, say to the right,  along a one-dimensional chain of length 
$L$ with periodic boundary conditions. Particles hop one site
to the right at rate $p_1$, if this site is not occupied. If a
particle has two empty sites in front of it, it may also move
two sites to the right with rate $p_2$. The stochastic Hamiltonian
of model A has the form (\ref{eq:Hdefper}) with $r=3$.
The local operator $\hat{h}_{k,k+1,k+2}$ acts on sites $k$, $k+1$ and
$k+2$ and its explicit form is given by the dynamics of the model. 

Model A is of obvious relevance for the modelling of traffic flow. 
It can be interpreted as a model for cars which have a maximum 
velocity $v_{max}=2$ moving on a single-lane highway. Up to now
no exact solutions of probabilistic traffic flow models with 
$v_{max}>1$ are known.

For the stationary state we make an ansatz of the form (\ref{eq:stat_period})
with $\mathcal{D}=(E,D)^t$ where $E,D\in \operatorname{End}(V)$ and
$V$ a vector space. Note that this corresponds to a grand canonical
description since (\ref{eq:stat_period}) is a superposition of states
with different particle numbers. However, in our case it is not difficult
to obtain results for fixed particle densities.

\begin{figure}[ht]
\begin{tabular}{ccc}
Flow $J$ & & \\
& \includegraphics[height=4cm]{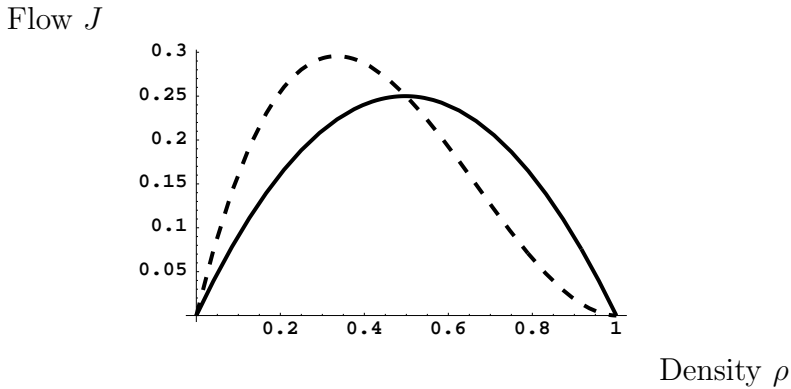}&\\
& & Density $\rho$
\end{tabular}
\caption{Fundamental diagram for model A with $p_2=1$: 
The dashed line corresponds to $p_1=0.01$ and the solid line to $p_1=0.99$.}
\label{fig:modela-fund}
\end{figure}

Using Proposition A from above, $|P_L\rangle_0$ is a stationary state
if one can find 6 operators $E$, $D$, $X_{11}$, $X_{12}$, $X_{21}$ and
$X_{22}$ such that they fulfil  the following relations,
given by the use of our cancelling mechanism
\begin{equation}\label{eq:modela-bulk}
\begin{aligned}
0&=X_{11}E-EX_{11},\qquad\\
p_1DEE&=X_{12}E-EX_{21},\qquad\\
-(p_1+p_2)DEE&=X_{21}E-DX_{11},\qquad\\
0&=X_{22}E-DX_{21},\qquad
\end{aligned}
\begin{aligned}
 p_2DEE&=X_{11}D-EX_{12},\\
 p_1DED &=X_{12}D-EX_{22},\\
-p_1DED &=X_{21}D-DX_{12},\\
 0 &=X_{22}D-DX_{22}.
\end{aligned}
\end{equation}
The algebra given by (\ref{eq:modela-bulk}) has the following  one-dimensional
representation with $D,E\in\mathbb{R}$:
\begin{equation}\label{eq:1d_repr}
\begin{aligned}
D&=1-E,\qquad\\
X_{11}&=xE/D+p_2E^2,\qquad\\
X_{21}&=x-p_1 ED,\qquad
\end{aligned}
\begin{aligned}
\ & {\phantom{strunk}}\\
X_{12}&=x\in \mathbb{R} \qquad \text{ (free parameter)},\\
X_{22}&=xD/E-p_1D^2.
\end{aligned}
\end{equation}
As a consequence the stationary probability distribution is a simple
product measure. The particle density $\rho$ is equal to $D$ and the
flow $J$ is simply given by
\begin{equation}
J(\rho,p_1,p_2)=\rho(1-\rho)[p_1+2p_2(1-\rho)],
\end{equation}
which gives us the fundamental diagram, i.e. the functional relation
between flow and density, at hand (see figure \ref{fig:modela-fund}).
Since model A does not exhibit a particle-hole symmetry, the 
fundamental diagrams are not symmetric with respect to $\rho=1/2$.

We have also investigated {\em model A} with parallel dynamics, which is
more appropriate for traffic models \cite{nagel-schreckenberg}.
Comparing the results of a so-called car-oriented mean-field theory (COMF)
\cite{COMF} with Monte Carlo simulations indicates that COMF gives actually
the exact fundamental diagram \cite{spaeta}.

As a next step we examined model A with open boundaries,
i.e.\ injection of particles at the left end of a chain of length $L$,
and removal of particles at the right end. For the left end we have
chosen the following input rates
\begin{equation}
\xymatrix{
|EE\cdots \ar[rr]^{\alpha_1}
\ar[drr]_{\alpha_2} & & |DE\cdots\\ 
& &  |ED\cdots\\
|ED\cdots \ar[rr]^{\alpha_3} & & |DD\cdots}
\end{equation}
The symbol $E$ denotes an empty site and $D$ an occupied one.
\newline
For the right end we have chosen the output rates
\begin{equation}
\xymatrix{
\cdots ED| \ar[rr]^{\beta_1} & & \cdots EE|\\ 
\cdots DE| \ar[rr]^{\beta_2} \ar[drr]_{\beta_3}& & \cdots ED|\\
& & \cdots EE| \\
\cdots DD| \ar[rr]^{\beta_4} & & \cdots DE|}
\end{equation}
Using the cancelling mechanism one gets the following relations, which
have to be fulfilled in addition to  (\ref{eq:modela-bulk})
\begin{equation}\label{eq:modela-boundary}
\begin{aligned}
-(\alpha_1+\alpha_2)EE&=-X_{11},\qquad\\
\alpha_2EE-\alpha_3ED&=-X_{12},\qquad\\
\alpha_1EE&=-X_{21},\qquad\\
\alpha_3ED&=-X_{22},\qquad
\end{aligned}
\begin{aligned}
\beta_1ED+\beta_3ED&=X_{11},\\
(\beta_2-\beta_1)ED&=X_{12},\\
-(\beta_3+\beta_2)ED+\beta_4DD&=X_{21},\\
-\beta_4DD&=X_{22}.
\end{aligned}
\end{equation}
We found that the one-dimensional solution (\ref{eq:1d_repr}) serves as a 
solution of (\ref{eq:modela-boundary}), if the boundary rates are given by
\begin{equation}\label{eq:modela-sol}
\begin{aligned}
\alpha_1&=p_1D+p_2DE,\qquad\\
\alpha_2&=p_2D,\qquad\\
\alpha_3&=\alpha_1,\qquad\\
\end{aligned}
\begin{aligned}
\beta_1&=-\beta_3 + p_1E + p_2E(1+E),\\
\beta_2&=-\beta_3 + p_1 + p_2E,\\
\beta_4&=p_1 E + p_2E^{2},\\
\end{aligned}
\end{equation}
where $\beta_3 \in\mathbb{R}$ is a free parameter, and
\begin{equation}
x=p_1D^2E-p_2DE^3.
\end{equation}

A more complete investigation of the properties of this model for general
values of the interaction parameters will be presented in \cite{spaeta}.

The second model we want to present (\emph{model B}) has again 
three-site interactions in the bulk. The same model has been
studied independently in \cite{schuetz} where equivalent results for
periodic and open systems have been found.
Similar to model A particles move along a chain of length $L$ exclusively 
in one direction. If a particle has two empty sites in front of it, it moves
one site with rate $1$; if only the next site is empty, the particle
performs the same move  with rate $\lambda $. 
For $\lambda < 1$ this model may be considered as a traffic flow model
with a so-called 'slow-to-start rule' (see \cite{barlo} and references
therein).

The stochastic process obtained from model B after a particle-hole
transformation~\footnote {And a parity transformation so that the
 particles again move from left to right.} is also very interesting
\cite{spaeta}.  Here the hopping probability depends on the occupation
number of the site directly behind the particle. If the site to the left
is occupied a particle moves to the right with rate $p_2$, if it is
empty it moves with rate $p_1$.  For $p_1 > p_2$ the particles prefer
to stick together which can lead to interesting clustering properties.
This model might have applications for granular matter and flocking
behaviour where similar
interactions have been studied (see e.g.\ \cite{haya}).

The matrix-product ansatz is the same as for model A and we find for
the bulk-algebra:
\begin{equation}
\label{eq:modelb-bulk}
\begin{aligned}
X_{11}E&=EX_{11},  \qquad \\
DEE&=X_{12}E-EX_{21},  \qquad \\
-DEE &=X_{21}E-DX_{11},  \qquad\\ 
X_{22}E&=DX_{21}, &  \qquad
\end{aligned}
\begin{aligned}
X_{11}D&=EX_{12},\\
 \lambda DED&=X_{12}D-EX_{22},\\
 -\lambda DED &=X_{21}D-DX_{12},\\
 X_{22}D&=DX_{22}.
\end{aligned}
\end{equation} 
For this algebra we found a two-dimensional representation which has
a structure generic for a 2-cluster approximation \cite{nagel-schreckenberg}
\begin{equation}
E=\begin{bmatrix} e & 1 \\ 0 & 0\end{bmatrix},\qquad
D= \begin{bmatrix} d &0  \\ 1 & 0\end{bmatrix},\qquad
ed=\frac{\lambda}{1-\lambda}
\label{eq:mpa_clust}
\end{equation}
\begin{equation}\label{eq_Xdef}
\begin{aligned}
X_{11}&=\frac{e}{d}(d+x)E, \qquad\\
X_{21}&=\frac{x}{d}DE, 
\end{aligned}
\begin{aligned}
X_{12} &= eD+\frac{x}{d}ED,\\
X_{22}&=xD.
\end{aligned}
\end{equation}

\begin{figure}[ht]
\begin{tabular}{ccc}
 Flow $J$ & & \\
& \includegraphics[height=4.5cm]{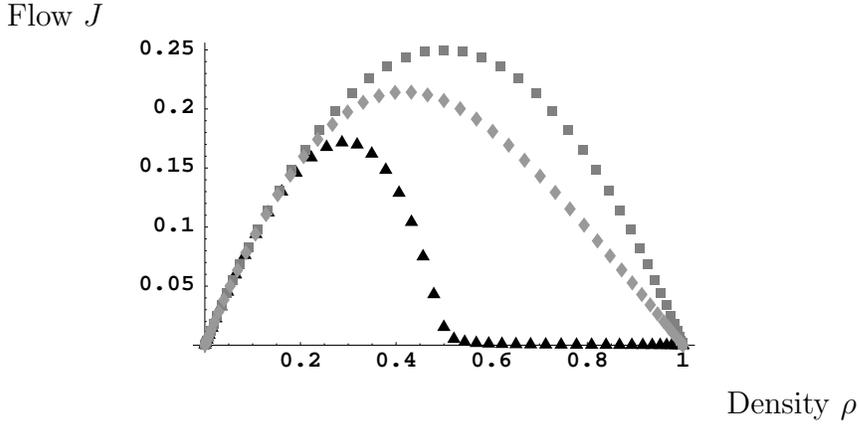}&\\
& & Density $\rho$
\end{tabular}
\caption{Fundamental diagram of model B: squares correspond to
  $\lambda=0.999$, triangles to $\lambda=0.001$ and diamonds to
  $\lambda=0.5$ .}
\label{fig:modelb-fund1}
\end{figure}

\begin{remark}
(i) It is not hard to prove, that any two operators $E$ and $D$, which
fulfil
\begin{equation}
\begin{split}
\label{eq:modelb-replace}
E^2&=eE,\\
D^2&=dD,\\
{\lambda}DED&={ed}D,
\end{split}
\end{equation}
with $e$, $d$ being c-numbers and the $X_{ij}$ defined in terms of
$E$, $D$ as in (\ref{eq_Xdef}), yield a representation of the algebra
(\ref{eq:modelb-bulk}).
\newline
(ii) A first consequence is that any representation of the Temperly-Lieb
algebra $T_3(\pm \sqrt{\lambda})$ gives a representation of
(\ref{eq:modelb-bulk}). This can be seen by defining $E=e_1$ and
$D=pe_2$ where $e_1$ and $e_2$ are the generators of $T_3(\pm
\sqrt{\lambda})$ and $p$ is a free parameter.
\end{remark}
With the use of the two-dimensional representation (\ref{eq:mpa_clust}) 
it is possible to calculate the expectation value of any observable in 
the stationary state (see e.g.\ \cite{derrida}). The fundamental diagram 
is given by
\begin{equation}
J(\rho)=\rho\left[1-\frac{1-\sqrt{1-4(1-\lambda){\rho}
(1-{\rho})}}{2(1-\lambda)(1-\rho)}\right]
.
\end{equation}
As an example we present some fundamental diagrams in 
figure \ref{fig:modelb-fund1}.
Again it is non-symmetric due to the lack of a particle-hole symmetry.

As for model A, we studied model B with open boundaries, having input
rates
\begin{equation}
\xymatrix{
|EE\cdots \ar[rr]^{\alpha_1} & & |DE\cdots\\ 
|ED\cdots \ar[rr]^{\alpha_2} & & |DD\cdots},
\end{equation}
and output rates
\begin{equation}
\xymatrix{
\cdots DE| \ar[rr]^{\beta_1} & & \cdots ED|\\ 
\cdots ED| \ar[rr]^{\beta_2} & & \cdots EE|\\
\cdots DD| \ar[rr]^{\beta_3} & & \cdots DE|}.
\end{equation}
According to our cancelling-mechanism this gives the following
relations, which supplement the algebra (\ref{eq:modelb-bulk}):
\begin{alignat}{1}
& \langle W| \begin{pmatrix} -\alpha_1EE \\ -\alpha_2 ED \\ \alpha_1
  EE\\\alpha_2ED\end{pmatrix}  =- \langle W| \begin{pmatrix}
  X_{11}\\X_{12}\\X_{21}\\X_{22}\end{pmatrix}\label{eq:modelb-bound1}\\
& \begin{pmatrix} \beta_2 ED \\ \beta_1DE-\beta_2ED \\
  \beta_3DD-\beta_1DE\\ -\beta_3DD \end{pmatrix} |V\rangle= 
\begin{pmatrix}X_{11}\\X_{12}\\X_{21}\\X_{22}\end{pmatrix}|V\rangle
\label{eq:modelb-bound2}.
\end{alignat} 
Again we found that the two-dimensional representation 
(\ref{eq:mpa_clust}), (\ref{eq_Xdef}) is also a representation for
(\ref{eq:modelb-bound1}), (\ref{eq:modelb-bound2}) if $x$ is given by
\begin{equation}
x_{\pm}=\frac{\lambda}{1-\lambda}\left[e+d\pm\sqrt{(e-d)^2+
  \frac{4}{\lambda}ed}\right],
\end{equation}
where $x_+$ and $x_-$ correspond to $\lambda > 1$ and $\lambda < 1$,
respectively, and the boundary rates satisfy
\begin{equation}
\begin{aligned}
\alpha_1 &= 1+\frac{x}{d}, \qquad \\
\beta_2&=\beta_3=-\frac{x}{d},\qquad\\
\langle W|&=\left(-\frac{ed}{x},1\right), \qquad
\end{aligned}
\begin{aligned}
\alpha_2&=\lambda\alpha_1,\\
\beta_1&=\frac{-xe}{d(e+x)},\\
|V\rangle&=\left(1,-\frac{de}{x}-d-e\right)^t.
\end{aligned}
\end{equation}
That means that we found a line in the parameter space
along which the stationary state can be written as
a matrix-product state with 2-dimensional matrices. The density
profiles calculated along this line are flat.
\newline



\section{Conclusion}
The generalization of the proposition of KS to stochastic processes
with interaction range $r\geq 2$ leads to the most general
cancelling-mechanism for such systems. The identification of this
mechanism is one of the main results of this letter.
Using this mechanism the existence of MPS for the stationary state of 
systems with boundary interaction is shown. As an application we were 
able to solve two models with three-site interactions.

These models are interesting by themselves since they might be of
relevance for the description of traffic flow or granular matter. 
We were able to find the stationary state of the periodic system
which in both cases is given by a finite-dimensional representation
of the algebra obtained from the MPA. 

For the open system with boundary interactions we could extend the
solutions of the periodic systems for special values of the
input and output rates. For general values of the boundary rates
one probably needs infinite-dimensional representations.

Both models have a fundamental diagram with only one maximum. Using the
argumentation of \cite{kolomeisky} one can expect that the phase
diagrams of model A and B with boundary interactions essentially look like 
the well known phase diagram of the TASEP \cite{derrida,domany-schuetz}. 

A generalization of the KS-proposition to other update schemes \cite{rsss}
has been presented in \cite{rajschreck}. A similar generalization is also 
possible in our case and will be presented elsewhere \cite{spaeta}.

\vfill\eject
\hyphenation{Sonder-forschungs-bereich}
{\bf Acknowledgements}\\[0.6cm]
 This work was performed within the research program of
the Sonderforschungsbereich 341, K\"oln-Aachen-J\"ulich.
We would like to thank K.~Krebs, H.~Niggemann, N.~Rajewsky,
V.~Rittenberg, L.~Santen and G.~M.~Sch\"utz for useful talks and
discussions.

\vfill
\eject
\newpage


\vfill\eject

\end{document}